\documentclass[12pt]{article}

\usepackage{epsf}
\usepackage{latexsym,euscript}
\usepackage[dvips]{graphicx}
\usepackage{amsmath}
\usepackage{amsfonts}
\usepackage{amssymb, epsfig}
\usepackage{xcolor}

\DeclareMathOperator{\arcsinh}{arcsinh}

\usepackage{color,soul}

\begin{document}

\begin{center}
{\Large \textbf{   Polyakov loop model with exact static quark determinant in the 't Hooft-Veneziano  limit: U(N) case
}}

\vspace*{0.6cm}
\textbf{S.~Voloshyn${}^{\rm a}$\footnote{email: billy.sunburn@gmail.com, s.voloshyn@bitp.kyiv.ua}}

\vspace*{0.3cm}
{\large \textit{${}^{\rm a}$ Bogolyubov Institute for Theoretical
Physics, National Academy of Sciences of Ukraine, 03143 Kyiv, Ukraine}}
\end{center}

\begin{abstract}
I investigate a $d$-dimensional $U(N)$ Polyakov loop model that includes the exact static determinant with $N_f$ degenerate quark flavor and depends explicitly on the quark mass and chemical potential. In the large $N, N_f$ limit mean field gives the exact solution, and the core of the  Polyakov loop model is reduced to a deformed unitary matrix model, which I solve exactly. I compute the free energy, the expectation value of the Polyakov loop, and the quark condensate. The phase diagram of the model and the type of phase transition is investigated and shows it depends on the ratio $\kappa =N_f/N$.
\end{abstract}

\section{Introduction}

In this paper, we focus on a 3-dimensional effective Polyakov loop (PL) model, describing the $(d + 1)$-dimensional $U (N)$ LGT with $N_f$ flavor of staggered fermions at finite baryon density.

The resulting PL  model we work with takes the form
\begin{eqnarray}
Z_{\Lambda}(N,N_f)   =  \!  \int \prod_x dU(x)
\exp \!  \left [ \beta\sum_{x,\nu} \ {\rm {Re}}{\rm {Tr}}U(x){\rm {Tr}}U^{\dagger}
  (x+e_\nu) \right ]  \! \prod_{f=1}^{N_f} B_q(m_f,\mu_f) \, ,
\label{sunpf}
\end{eqnarray}
where the determinant is taken over group indices.
The coupling  of  this effective model depends on the temporal gauge coupling and can be expressed as
\begin{eqnarray}
\beta \equiv D_{fund}(\beta_g) \ = \ \left ( \frac{C_{f}(\beta_g)}{s_{1}(I) C_{0}(\beta_g)} \right )^{N_t} \ , \
C_{f}(\beta) \ = \ \sum_{k=-\infty}^{\infty} \ {\rm det}\, I_{\delta_{i,1} - i + j + k}(\beta_g)_{1\leq i,j \leq N} \ .
\label{D_coeff}
\end{eqnarray}
In this model the matrices $U(x)$ play the role of Polyakov loops, the only
gauge-invariant operators surviving the integration over spatial gauge fields
and over quarks. The integration in~(\ref{sunpf}) is performed with respect to the Haar measure
on $G$.  

In case of exact static determinant with $N_f$ degenerated quark flavors  (or $1d$ QCD , see review of dual models \cite{Philipsen_19},\cite{pl_dual20})
\begin{equation}
\prod_{f=1}^{N_f} B_q(m_f,\mu_f) \ = \ B_{st}(h_+^f,h_-^f) \ = \
A_{st} \, {\rm det} \left [ 1 + h_+^f U(x) \right ]^{N_f} \, \left [ 1 + h_-^f U^{\dagger}(x) \right ]^{N_f} \ ,
\label{Zf_stag}
\end{equation}
where
\begin{equation}
A_{st}  = h^{-N N_f} \ , \ \ \ h_{\pm} = h e^{\pm \mu} \ , \ \ \
h = e^{-N_t \arcsinh m}  \ .
\label{hpm_stag}
\end{equation}
An action of this type describes the interaction between Polyakov loops ${\rm {Tr}}U(x)$ at finite temperature and in the presence of $N_f$ heavy degenerate quark flavors. The parameter $h$ can be related to the mass of the quark (its exact form depends on the kind of lattice fermion) and $\mu=\beta\mu_q$, where $\mu_q$ is the chemical potential of the quark. 

We solve this model in the 't~Hooft-Veneziano limit \cite{Hooft_74}, \cite{Veneziano_76}:
\begin{equation}
N \to \infty, \quad N_f \to \infty, \quad \kappa = \frac{N_f}{N} = \text{const},
\end{equation}
where the mean-field approximation becomes exact (in the spirit of \cite{damgaard_patkos}, \cite{christensen12}). The central analytical object in this limit is a deformed unitary matrix model:
$$
 A_{st} \int  dU  e^{N  \left ( g_+ {\rm {Tr}}U +  g_- {\rm {Tr}}U^{\dagger} \right)}   \det  \left[1+ h_+ U \right]^{N_f} \left[1 +  h_- U^{\dagger}  \right]^{N_f} 
$$
$$
 \approx A_{st} \int dU  e^{N  \left ( (g_++ \kappa h_+) {\rm {Tr}}U +  (g_- + \kappa  h_-) {\rm {Tr}}U^{\dagger} \right)} 
$$
which generalizes the classical unitary matrix model of GWW \cite{gross_witten}-\cite{wadia}.
So far in the context of the PL model, the static fermion determinant was investigated in the heavy quark approximation  \cite{{pisarski18}}, \cite{pl_largeN_conf21},  \cite{pl_largeN21} .

Our main technical task was to solve this model in the 't Hooft-Veneziano limit. We solve this matrix model exactly at large $N$ and derive analytic expressions for the free energy, the Polyakov loop expectation value, and the baryon density. We determine the whole  phase diagram and identify the presence of a third-order phase transition and switch to a first-order one at $\kappa \to 0$. This result is a generalization of the clasical GWW solution \cite{gross_witten},\cite{wadia}.
Case $h=1$ considered in \cite{Russo_2020}-\cite{SANTILLI_2020}, $\beta  =0$ in \cite{1D_QCD}. 

This paper is organized as follows. In Sec.~2 we derive all the additional analytical work including the mean field approach for a large $N$,  representation of $U(N)$ unitary matrix model (which is the core of this PL model), and present the exact solution of this matrix model in the limit of large $N, N_f$. In Sec 3 we apply a solution of the matrix model to obtain phase diagram of PL model,  expressions for its free energy, and observables like average PL and correlation function.  The summary is presented in the final section.

\section{ PL model and one-site integral }

\subsection{Mean field method for large N  in $d$ dimension}

The partition function of the $U(N)$ and $SU(N)$ PL model with local nearest–neighbor interaction can be rewritten in the large–$N, N_f$ limit as  (see \cite{christensen12})
\begin{eqnarray}
\label{PF_spindef}
Z_{\Lambda}(N,N_f)  =
\int \ \prod_x \ dU(x)
 \exp \left[ \beta \ \sum_{x,\mu} \ {\rm Re}{\rm Tr}U(x) {\rm Tr}U^{\dagger}(x+e_{\mu}) \right] A_{st}\\
\prod_x \det\left[1+ h_+ U(x)\right]^{N_f} \left[1 +  h_- U^{\dagger}(x)  \right]^{N_f}
=
\left[
e^{ -  N N_f \frac{\beta d}{\kappa}  \,  {\rm Re}\,{\rm ww}^{\dagger}}
\sum_{q = - \infty }^{\infty}  \Xi_q (g_{\pm},h_{\pm})
\right]^{L^d}
\nonumber
\end{eqnarray}

In the joint limit of large–$N$, large–$N_f$, the mean field approach becomes exact, provided that the ratio $N_f/N = \kappa$ is kept finite. The sum over $q$ is then replaced by an integral under $q \to u / N_f$. The resulting core of the Polyakov loop model with the exact static fermion determinant takes the form of a deformed unitary matrix model 
\begin{eqnarray}
 \sum_{q = - \infty }^{\infty} \Xi_q (g_{\pm},h_{\pm}) = A_{st} \!
 \int  dU \
 e^{N  \left ( g_+ {\rm Tr}U +  g_- {\rm Tr}U^{\dagger} \right)}
 \det\left[1+ h_+ U \right]^{N_f}
 \left[1 +  h_- U^{\dagger}  \right]^{N_f}  =  \label{defmatrxmodel} \\
 =
 e^{  N N_f  F_{SU(N)}}  =
 \int_{- \infty }^{\infty} d u \, e^{ N N_f S_{eff}[g_{\pm}, h_{\pm}; u]} \ , 
\end{eqnarray}
where we denote
\begin{equation}\label{crit_w}
    \textrm{w} \, = \,   \left\langle  \frac{1}{N} {\rm {Tr}} U  \right\rangle   ,  \ \textrm{w}^{\dagger} \, = \   \left\langle  \frac{1}{N} {\rm {Tr}} U^{\dagger} \right\rangle   , \  g_+ \, = \, b  \textrm{w}^{\dagger}   ,  \  \ g_- \, = \,  b \textrm{w} \, , 
\end{equation}
and  $b= \beta d$.

In the large $N,N_f$ limit, the mean field approach becomes exact:
\begin{eqnarray}
&&Z_{\Lambda}(N,N_f) = [e^{ N N_f( -\frac{ g_+ g_-}{\kappa b} + F_{SU(N)}) } ]^{L^d } =  [e^{   N N_f  f_{SU(N)}}  ]^{L^d } \, ,
\end{eqnarray}

where we introduce $S_{eff} \equiv S$ 

The mean–field equations (with $b = \beta d$) read
\begin{eqnarray}\label{crtwi}
    \frac{g_-}{ \kappa b}
    =  
    \frac{\partial}{\partial g_+}  S_{eff}[\kappa, g_{\pm}, h_{\pm}; u]
    ,
    \qquad
    \frac{g_+}{\kappa b}
    =
    \frac{\partial}{\partial g_-}  S_{eff}[\kappa, g_{\pm}, h_{\pm}; u]  \, ,
    \label{meaneq}
\end{eqnarray}
together with an additional condition reflecting the $SU(N)$ nature of the model (the extremum equation in $u$):
\begin{eqnarray}\label{ctytyw}
     \frac{\partial}{\partial u}  S_{eff}[\kappa, g_{\pm}, h_{\pm}; u]
     =
     0
     \label{u_eq}
\end{eqnarray}

\subsection{Formulation of deformed unitary matrix model}

The variables $g_+$ and $g_-$ play a technical role and attain their physical values only after embedding the effective one–site model into the full $d$–dimensional PL model. At this stage it is useful to redefine
$g_{\pm} \to g_{\pm} e^{\mp \mu}$ so that $h_\pm = h$, although we keep the notation $h_\pm$ whenever needed.

We can reformulate (\ref{defmatrxmodel}) with the help of parametrization of the group matrix $U$ using its eigenvalues as 
\begin{eqnarray}
  && \Xi (g_{\pm},h_{\pm}) =  A_{st} \prod_{n=1}^N    \sum_{k_n, l_n=0}^{N_f} h_+^{k_n} h_-^{l_n} \binom{N_f}{k_n} \binom{N_f}{l_n}  \sum_{r_n , s_n =0 }^{\infty } \frac{(N g_+)^{r_n}}{r_n!}
 \frac{(N g_-)^{s_n}}{s_n!} \ Q
 \end{eqnarray}
where
$$
Q= \int_{SU (N)} d U \prod_{n= 1}^N e^{i (k_n - l_n+ r_n- s_n) \phi_n } = \sum_{q = - \infty }^{\infty} \det_{1 \leq i , j \leq N} \delta_{i -j + k_i - l_j + r_i - s_j+ q,0}
$$
and with help of determinant (that linked with skew  Schur function, see  \cite{sun_int}):
 \begin{eqnarray}
 && \Xi (g_{\pm},h_{\pm}) =  A_{st} \sum_{q = - \infty }^{\infty} \prod_{n=1}^N \sum_{r_n , s_n =0 }^{\infty } \frac{(N g_+)^{r_n}}{r_n!} \frac{(N g_-)^{s_n}}{s_n!}   \sum_{l_n =0}^{N_f} h_+^{l_n -r_n+ s_n + q} h_-^{ l_n} \nonumber \\
 && \times \binom{ N_f}{l_n} \det_{1 \leq i , j \leq N}  \binom{ N_f}{-i +j + l_i -r_i+ s_j + q} \, .
 \label{det_rep}
\end{eqnarray}
All the details needed and another formulation will be found in the Appendix \ref{formulation}.

\subsection{Solution of $U(N)$ deformed unitary matrix  model and its phase diagram }

All calculations were performed as calculations of the corresponding determinants (see details in \ref{Shur_func}) with generalization of particular cases for finite $N$. All these formulae work for $SU(N)$, so we take $q=0$ to get the case $U(N)$ ($S_{eff} [\kappa, g_{\pm}, h_{\pm}; 0] \equiv F $).  After exponentiation of the series, we take the 't Hooft-Veneziano limit and obtain a strict expansion of the free energy as small $h$ series that have only three different terms (up to $g^2$ as we see in (\ref{F1})) and small combined $m$ and $g_{\pm}$ series that contain infinite many terms (\ref{F2U(N)}) (see more details in \ref{Strict}). 

After some algebra, we have the next exact expression for the free energy $U(N)$ matrix model in the case $g_+=g_-=g$ (and we denote $F_{U(N)}\equiv F_{1,2}[\kappa, g_{\pm}, h_{\pm}]$ depending on the region) :
\begin{eqnarray}
F_1 [\kappa,  g, h] = - \log h   - \kappa \log(1- h^2) +2  g h + \frac{g^2}{ \kappa} \,  , \label{F1} \\
F_2 [\kappa,  g, h] =\frac{2 g }{ \kappa } +\frac{\overline{g} \left(2 \overline{g}^2+2 (\overline{g}+1) \kappa-5 \overline{g}+2\right)}{\kappa(2 \overline{g}-2 \kappa-1)}-\frac{(\kappa+1)^2 \log (1-2 \overline{g}+ \kappa)}{\kappa }+  \\
   + \frac{\log (1-2 \overline{g})}{2  \kappa}+\frac{(2 \kappa+1)^2 \log (1-2 \overline{g}+2 \kappa)}{2 \kappa}+\frac{(2 \kappa+1) \log   \left(\cosh ^2\left(\frac{m}{2}\right)\right)}{2 k}- \kappa \log 4 \kappa \nonumber  
\end{eqnarray}
where
\begin{eqnarray}
\overline{g} \equiv
 \overline{g} (G, \kappa)= \frac{1}{3}   (1+G+ \kappa)-\frac{(1-2 G+ \kappa)^2}{6 \sqrt[3]{P}}-\frac{\sqrt[3]{P}}{6} \label{solUN}
\\
P= -8 G^3+12 G^2 (\kappa+1) +3 G (7 \kappa^2 - 4 \kappa-2)+(\kappa+1)^3 + \nonumber
\\   
   +3  \kappa \sqrt{ 3 G \left(-16 G^3+24 G^2  ( \kappa+1)+3 G ( \kappa-2) (5  \kappa +2)+2 ( \kappa+1)^3\right)} \nonumber
\end{eqnarray}
where $\xi= 2 \kappa +1$ and $G = g \cosh^2\left(\frac{m}{2}\right)$ .

This solution relies on the property that, almost everywhere in the
series expansion, the variables $m$ and $g$ appear only through the
combination $G= g \cosh^2 \frac{m}{2}$ as can be seen in (\ref{F2U(N)}). This property allows one to bypass the need to use the method of orthogonal polynomials \cite{orthog_polynom}. We apply a method of reconstruction that is kind of algebraic lifting from the critical line. The reconstruction proceeds as follows.  First, we need to know the critical line exactly (see (\ref{U(N)critline}) - which is  determined as a series in
$g$ and then resummed to the exact expression $h_{cr} = \frac{1-2 g}{\xi -2 g}$.
Second, consider the following property: both free energies become identical on the critical line:
\begin{eqnarray}
\label{critline_eqn}
F_2[\kappa,  \overline{g},  \frac{1-2 \overline{g}}{\xi -2 \overline{g}}] =F_1[\kappa, \overline{g},  \frac{1-2 \overline{g}}{\xi -2 \overline{g}}] \, .
\end{eqnarray}
Thus, the equality on the critical line provides the exact value of the
$F_2$ restricted to the phase boundary (because  we know $F_1$ exactly). We will see below that $\overline{g}= \overline{g} (G, \kappa)$ is the solution of a cubic algebraic equation. Third, if we want to reconstruct the true dependence of $F_2$, we make use of the fact that the property $F_2[\kappa, g, h]= F_2[\kappa, G]$ is almost true. That means that the next identity holds after projection:
\begin{eqnarray}
\label{properyU(N)}
G=\overline{g} \cosh ^2\left(\frac{m}{2}\right)|_{m \to - \log[ \frac{1-2 \overline{g}}{\xi -2 \overline{g}}]}=\frac{\overline{g} ( 2\overline{g}-\kappa -1)^2}{(2 \overline{g}-1) (2 \overline{g}-2 \kappa -1)} = g \cosh^2 \frac{m}{2}
\end{eqnarray}
This relation provides the algebraic map between the invariant
combination $G$ and the parameter $g$ on the critical line.
This holds for all sides of equality (however $F_1[\kappa, g, h]= F_1[\kappa, G]$ is not true at all). Therefore we only need to invert the relation (\ref{properyU(N)})
and express $\overline g$ as a function of $G$. Applying this substitution to $F_1$ yields the exact expression for $F_2$. The only obstruction to expressing the result entirely in terms of $G$ comes from two terms in the free energy which do not combine into the 
invariant combination $G$ that "spoil" the  property $F_2[\kappa, g, h]= F_2[\kappa, G]$ as we see in (\ref{F2U(N)}). These terms are therefore separated explicitly, 
while the remaining part depends only on $G$.

After removing these two non-invariant contributions, the remaining 
expression can be reconstructed by inverting the algebraic relation 
(\ref{properyU(N)}) and finally restoring these two terms in the full expression for the free energy
\begin{eqnarray}
\label{equality}
F_2[\kappa,  g,  h ] =F_1 \left[\kappa, \overline{g} (G,\kappa),  \frac{1-2 \overline{g} (G,\kappa) }{\xi -2 \overline{g} (G,\kappa)} \right]-\frac{2 \overline{g} (G,\kappa)}{\kappa } - \nonumber \\ 
  - \frac{(2 \kappa +1)  }{2 \kappa }  \frac{(2 \overline{g} (G,\kappa) - \kappa -1)^2}{(2 \overline{g} (G,\kappa) -1) (2 \overline{g} (G,\kappa) -2 \kappa -1)}  +\frac{2 g}{\kappa }+ \frac{(2 \kappa +1) \log \left(\cosh^2 \frac{m}{2} \right)}{2 \kappa }  \, .
\end{eqnarray}

The  critical line is defined by 
\begin{eqnarray}
\label{U(N)critline}
  h_{cr} = \frac{1-2 g}{\xi -2 g} \ , or \, m =  \mp \log[\frac{1-2 g}{\xi -2 g}] \, .
\end{eqnarray}
that provides the 3rd order phase transition. The third derivative of $\Delta F=F_2[\kappa,  g,h]- F_1[\kappa,  g, h]$ at  the phase transition exhibits the finite jump  
$$
\Delta F^{'''}_h|_{h=h_{cr}} 
= \frac{(1-2 g+2 \kappa)^6}{4 \kappa ^2 \left(2 \kappa^2 + 3 (1-2 g)^2 \kappa -(2 g-1)^3\right)} \, .
$$

\begin{figure}[htb]
\centering{ \includegraphics[scale=0.5]{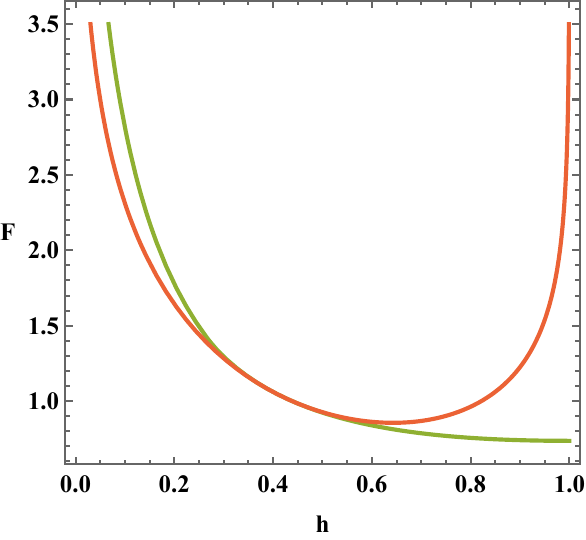}  \ \ \ \includegraphics[scale=0.6]{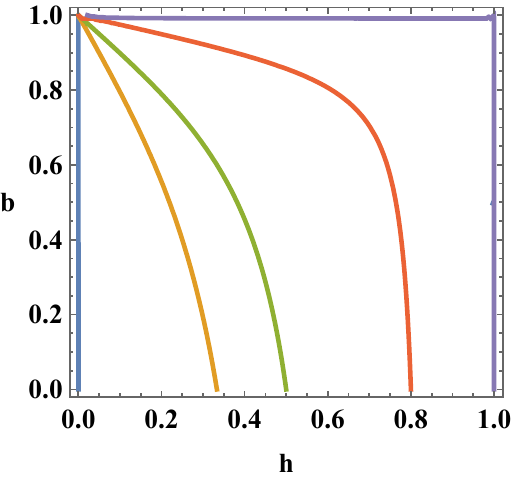} }
\caption{Free energies of the PL model as function of  $h$ at $\kappa=1/2$ and $b=5/12$ (right panel), red - $F_1$ , green - $F_2$ (left panel) and    phase diagram  of PL model in $h$-$b$ coordinates   with fixed  $\kappa$ = 0,  1/8, 1/2, 1, $\infty$ (corresponding colors counted from blue to violet )  (right panel).  
\label{fig1}}
\end{figure} 

\section{ Free energy of the PL model in mean field approximation}

The mean field equation for $F_1[g,h; 0]$ gives
\begin{eqnarray}
g_{MF} =  \frac{ b \kappa h}{1-  b } 
\end{eqnarray}

Free energy of the  confinement phase
\begin{eqnarray}
f_{U(N)}^1=    -  \log h +  b   \, \frac{\kappa h^2}{1-  b} - \kappa \log(1- h^2) 
\end{eqnarray}

For second free energy we arrive to next mean field equation: 
\begin{eqnarray}\label{mean_fiels}
     \frac{ g_{MF}}{ \kappa b } \ = \  \frac{1}{2} \frac{\partial}{\partial g }  F_{2}[\kappa, g, h] = 
        \frac{1}{g_{MF}}\frac{ \overline{g}\left(2 
     \overline{g}^2-3 \overline{g} +\kappa +1\right) }{ \kappa 
   (2 \overline{g}- \kappa -1)}+\frac{1}{\kappa }
\end{eqnarray}
where $\overline{g}=\overline{g}(G, \kappa)$ is as in  (\ref{solUN}) and $G = g_{MF} \cosh^2\left(\frac{m}{2}\right)$ .
Free energy of 
\begin{eqnarray}
f_{U(N)}^2= -\frac{ g_{MF}^2}{ \kappa  b} + F_2[\kappa, g_{MF}, h]
\end{eqnarray}

In all areas we have a third order phase transition except line $b =1$ where we have switching to the first order. In the limit of $\kappa \to 0$ type of phase transition also switch to the 1st order.  That`s all in the spirit of PL lattice model in the `t Hooft-Veneziano limit (which is solvable with the help of famous GWW model solution) but in our case we obtained a slightly more complicated expression for the critical line:
\begin{eqnarray}
\label{U(N)_crit_PL}
  h_{cr} (b, \xi) = -\frac{(b -3 )\xi +2+\sqrt{((b -3) \xi+2)^2+8 (b-1) (\xi-1)}}{4 (\xi-1)} \ , 
\end{eqnarray}

\begin{figure}[ht]
\centering{ \includegraphics[scale=0.87]{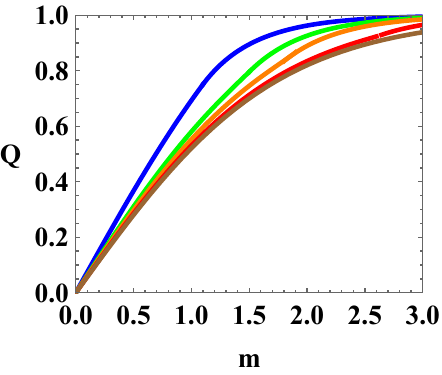}  \ \ \ \includegraphics[scale=0.36]{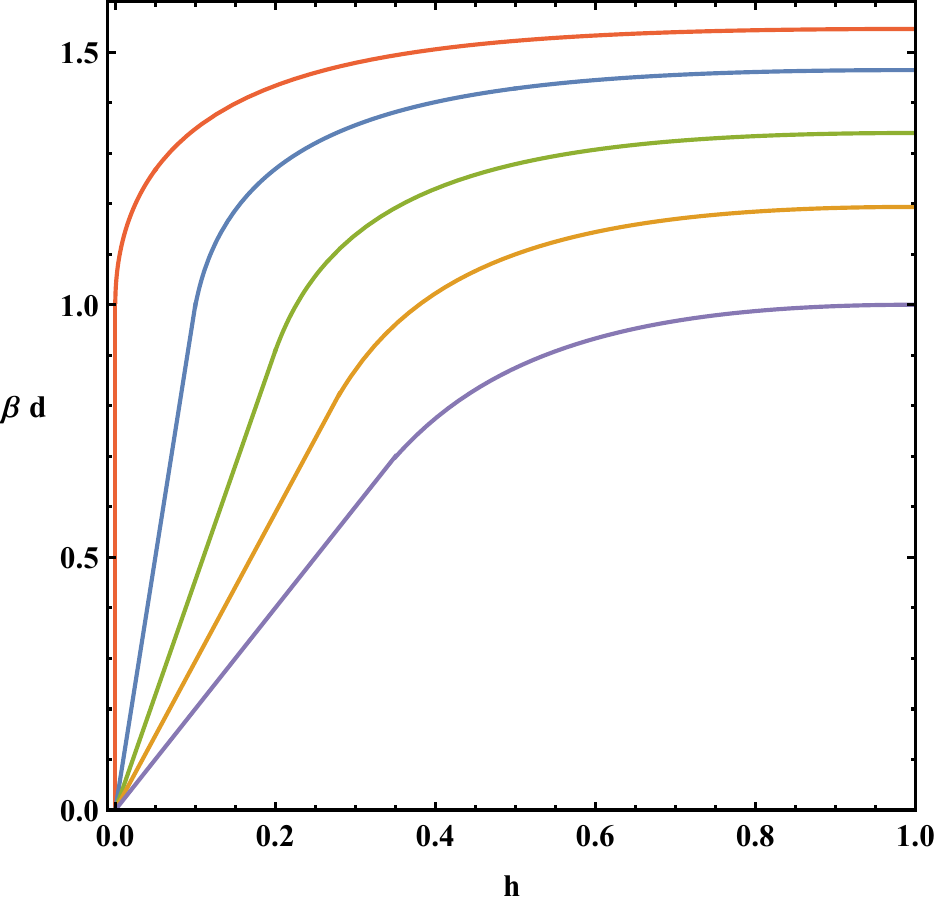} }
\caption{Quark condensate vs. $m$ $\kappa=1$ ($b$=0, 0.5, 0.66, 0.85, 1)  (left panel) and  the average  PL in in $h$-$b$ coordinates  at $\kappa=1$ ($b$=0, 0.36, 0.52, 0.8, 1)   (left panel) . Colors counted from top to bottom.
\label{fig2}}
\end{figure}

Average PL and quark condensate.

Quark condensate defined as:
$$
Q= \frac{\partial f_{U(N)}^{1,2}}{\partial m}
$$
Average PL: 
\begin{eqnarray}\label{crtwi}
    W(1,\mu)  \ = \  W^{\dagger}(1,-\mu)  \ = \  \begin{cases}
			\frac{  \kappa  h \, e^{-\mu}}{1-  b} \ , \
			& h < h_{cr}  \ , \\
		     \kappa  
          g_{MF}\, e^{-\mu} ,  & h_{cr} <  h < 1/h_{cr}    \ , \\
		\frac{  \kappa /h \, e^{-\mu}}{1-  b} \ , \
		& h > 1/h_{cr} \ ,
	\end{cases}
\end{eqnarray}
where $g_{MF}= g[b, \kappa, h]$ is the solution of eq. (\ref{mean_fiels}) .

Chiral symmetry is restored only at $m=0$ (See Fig. (\ref{fig2})).

\section{Summary and Perspectives}

The $U(N)$ PL model that takes into account exact static quark determinant with $N_f$ degenerate flavors of quarks investigated in the mean-field approximation.  First, we solved the deformed unitary matrix model. Second, apply this solution as the core that is needed to represent the mean field of the PL model.

Let us briefly summarize our main results: 
\begin{itemize}

\item 
We study  next the deformed unitary matrix model (that is  core  of the early mentioned PL lattice model )
\begin{eqnarray}
&& \Xi (g_{\pm},h_{\pm}) = A_{st} \int  dU  e^{N  \left ( g_+ {\rm {Tr}}U +  g_- {\rm {Tr}}U^{\dagger} \right)}   \det  \left[1+ h_+ U \right]^{N_f} \left[1 +  h_- U^{\dagger}  \right]^{N_f}  . \nonumber 
\end{eqnarray}
 We present a solution of such models in the 't Hooft-Veneziano limit \cite{Hooft_74,Veneziano_76}: $g\to 0, N\to\infty, N_f\to\infty$ such that the product $g^2 N$ and the ratio $N_f/N=\kappa$ are kept fixed ($g$ is the coupling constant).
 
\item 
Obtained expansion in small $h$ and small $m$.
We find an exact solution of this model for the $U(N)$ case  in  't Hooft-Veneziano limit and for  all $m$ and $g_+ = g_-$ I established 3rd order phase transition   with critical line  $h = \frac{1-2 g}{2 \kappa+1  -2 g}$ .

\item 
 The model around  $h=0$ reproduces the GWW solution in terms of $g$ \cite{gross_witten}, \cite{wadia}:
$$
 F_1= - \log h +\frac{1}{\kappa}(g^2) +2 g \, h + O(h^2)
$$
$$
 F_2=- \log h +\frac{1}{\kappa}( 2g - \frac{3}{4} - \frac{1}{2} \log 2 g ) + \frac{(4 g -1)  }{2 g} h +O(h^2)
$$
At $g =0$, the model reduces to a one-dimensional QCD \cite{1D_QCD}.

\item
Critical line of the general case $g_+ \neq g_-$ (see (\ref{full_crit})):
$$
 h = \frac{1 -  g_+ - g_-}{\xi -  g_+ - g_-}  + \frac{2 (g_+ - g_-)^2 \kappa^3}{(1 + k) (1 + 2 k)^3}\left(1 +\frac{ (g_+ + g_-) (8 \kappa +5) }{(\kappa +1) (2 \kappa +1)} + O (g^2) \right)\ .
$$
 That guaranties a third-order phase transition not only for $g_+=g_-$ case, but also for the general one. When we put $\kappa \to 0$ we switch the order of the phase transition to the first one. 
 
\item 
The solution of this matrix model could serve as the generation functional for the more sophisticated PL model that takes into account any powers of spin $({\rm {Tr}}U)^r$, and as the approach behind the mean field one.

\end{itemize}
The PL  model was investigated in the mean field approximation (in the limit large $N, N_f$ this approach becomes the exact):
\begin{itemize}

\item 
I obtained the free energy and the critical line given by (\ref{U(N)_crit_PL}).  It can be summarized briefly as follows.  The model in the limit of large  $N, N_f$ has two phases.  When $h=0$ (infinitely heavy quarks) one finds a 1st order phase transition.   It exhibits a third-order phase transition GWW type. 

\item 
  Limit $\kappa \to 0$ drive to domination of only one phase ("confinement") $F_1$ with switching of the phase transition to the first order. In the opposite limit $\kappa \to \infty$ survives the other phase ("deconfinement") without phase transition. 
 
\item 
We calculate some local observables, such as the average PL and the quark condensate.

\end{itemize}

As a natural extension of the present work, I plan to consider
the more involved $SU(N)$ case, which includes nonzero baryon density.
As is well known, the large-$N$ limits of $U(N)$ and $SU(N)$ theories
are not identical \cite{largeN_sun}, which manifests itself through the
emergence of baryon density. The appearance of nonzero baryon density
signals a phase transition at finite chemical potential and can be
viewed as a continuation of the $U(N)$ case (in simpler models usually
only one such critical line exists \cite{pl_largeN21}). In our model
with the exact quark determinant an additional critical line appears
when fermion saturation is reached; see also our discussion of 1d QCD
in Ref.~\cite{1D_QCD}.

\section*{Acknowledgements}
I thank to O. Borisenko and V. Chelnokov for stimulating discussions; this work was supported by the Simons Foundation (Grant SFI-PD-Ukraine-00014578).

\section{Appendix}

\subsection{Different representation of the deformed unitary matrix model}
\label{formulation}

Collection of all representations.

1) Small $g$ expansion around $1d$ QCD (each order is the corresponding averaged spins of $1d$ QCD as the Schur skew function):
\begin{eqnarray}
&& \sum_{q}  \Xi_q (g_{\pm},h_{\pm})= A_{st} \sum_{q} \ e^{ N q \mu} \ \sum_{r,s=0}\frac{ (N g_+)^r (N g_-)^s}{r! s!}  \sum_{l, l^{'}} h_+^l h_-^{l^{'}} \sum_{\nu \vdash r , \mu \vdash s } d(\nu) d(\mu ) \nonumber \\
&& \sum_{\sigma  \vdash r+ l} s_{\sigma /\nu}(1^{N_f}) s_{N^q \sigma/\mu }(1^{N_f}) \, \delta_{r+ l, s+ l^{'}+ q N} \end{eqnarray}

2) Small $m$ expansion .
Expanding in the most symmetrical way:
the partition function (\ref{det_rep}) can be reexpanded by small $m$ expansion .
\begin{eqnarray}
\sum_{q}  \Xi_q (g_{\pm},h_{\pm}) \ =  \ \sum_{q} \ e^{ N q \mu} \ Z_0^{N, N_f} (q) \sum_{r, s =0 }^{\infty } \frac{(N g_+)^{r}}{r!} \frac{(N g_-)^{s}}{s!} \sum_{n=0}^{\infty}A_{r,s}(q,n) m^{ 2 n}  \ . \label{PF_form_343}
\end{eqnarray}

General structure of the expansion near $h=1$
$$
\prod_{n = 1}^N e^{- l_n m }  = \prod_{n = 1}^N (1 -  l_n m + \frac{l_n^2}{2!} m^2 - \frac{ l_n^3}{3!} m^3 +...) = 1 - \sum_{k =1}^N l_k m + 
$$
$$
+ \frac{1}{2!} \left (  \sum_{k =1}^N l_k^2 + \sum_{k\neq r =1}^N  l_k l_r  \right) m^2 -  \frac{1}{3!} \left(  \sum_{k =1}^N l_k^3 + 3 \sum_{k\neq r  =1}^N l_k^2 l_r  + \sum_{k\neq r \neq s =1}^N  l_k l_r l_s   \right) m^3 + ... =
$$
$$
= 1+ \frac{1}{2!} T_1 m^2 + \frac{1}{4!} \left ( T_2 + \sum_{k\neq r =1}^N  T_1 T_1 \right) m^4 + O( m^6)
$$
We choose $l \to 2 l- N_f+a$ and $a= -r_i+ s_j -i+j+ q$ to make the most efficient way  of calculation. 
$$
Z = e^{-(2 l - N_f +a ) m} = 1 - (2 l - N_f+ a) m  + \frac{(2 l - N_f + a)^2}{2}  m^2 - \frac{ (2 l - N_f + a)^3}{3!} m^3 + ...
$$
Making summation (denoting )
$$\sum_{l=0}^{N_f} {N_f \choose l} {N_f \choose l+a}  = {2 N_f\choose N_f +a}= T_0,
$$
$$
\sum_{l=0}^{N_f} {N_f \choose l} {N_f \choose l+a} (2 l - N_f + a) = 0, 
$$
$$
\frac{1}{T_0} \sum_{l=0}^{N_f} {N_f \choose l} {N_f \choose l+a}\frac{(2 l - N_f + a )^2}{2!}  = -\frac{(a^2-N_f^2)}{2 (2N_f-1)} = T_1 , 
$$
$$\sum_{l=0}^{N_f} {N_f \choose l} {N_f \choose l+a} \frac{(2 l - N_f + a )^3}{3!}  = 0 ,
$$
$$
\frac{1}{T_0}\sum_{l=0}^{N_f} {N_f \choose l} {N_f \choose l+a}\frac{(2 l - N_f + a )^4}{4!}   = 
\frac{\left(a^2- n^2\right) \left(4 n +3 \left(a^2- n^2\right) \right)}{4!  (2 n -3) (2 n -1)}  = T_2 , 
$$
and so on.
Calculation of determinants:
$$
  \det_{1\le i, j \le N} \ T_0 =  \frac{ G[N+ 2 N_f+1] G[N+1] G[N_f-q+1] G[N_f+q+1] }{ G[2 N_f+1] G[N+ N_f- q+1] G[N+ N_f+ q+1] } = Z_0^{N, N_f} (q)   \ .
$$
where $G(x)$ is the Burness G function, and
$$
A_{0,0}(q,1)= \frac{1}{ Z_0^{N, N_f} (q) }\sum_{k=1}^N \det_{1\le i,j\le N} \left[ T_1^{i, j} \delta_{k, i}+ (1-\delta_{k, i}) \right] T_0^{i \, j} = \frac{ N (N + 2 N_f) (N_f^2 -q^2) }{2 (4 N_f^2-1)} \ ,
$$
and so on.
For calculation (as example) $A_{1,1}(q,n)$ we need in $A_{0,0}(q,n)$ make a  shift  $-i+j \to -i +\delta_{i,1}- \delta_{j,1} $ (as a "sources" dictated Young tableau) and, in general, $A_{r,s}(q,n)$ this is looks like a sum by all partition of  both - $r$ (associated with some  $\sum_p t_p\delta_{j,p}$) and  $s$ (associated with some  $\sum_p t^{\prime}_p \delta_{i,p}$) .

\subsection{ Small $h$ and $m$  expansion for arbitrary  $N, N_f$}
\label{Shur_func}

{\bf Small $h$ expansion }

The partition function reads 
\begin{eqnarray}
& \Xi (g_{\pm}, h\approx 0)=  A_{st} \, \sum\limits_{q=0}^{N_f} \,  e^{N q \mu} \, h_+^{N q} \ C_{N,N_f}(q) \biggl (Y_0+ N (\frac{g_+}{h} Y_{11}+  g_- h \, Y_{12})+ \nonumber \\
& +N^2(g_+^2/ h^2 Y_{21}+ g_- g_+ Y_{22}+g_-^2 h^2 Y_{23})+ O (g_{\pm}^3) \biggr )\ , 
\label{PF_form_4}
\end{eqnarray}
where
\begin{eqnarray}
&C_{N,N_f}(q) =  \underset{1\le k,l \le N}{\det} \ \binom{ N_f}{-k+l + q} = \frac{G(N+1) G(N+N_f+1) G(q+1) G(N_f+1-q)}{G(N_f+1) G(N+q+1) G(N+N_f+1-q)} \ , 
\label{CN_Nf_def} \\
&Y_0(q)= 1 + h^2 N N_f \ \frac{N_f - q}{N + q}+h^4 \frac{N N_f (N_f-q) \left[\left(N^2-1\right)
\left(N_f^2+1\right)-q  (N_f-N+q) (N N_f-1) \right]}{2 (N+q-1) (N+q) (N+q+1)} +O (h^6) \nonumber \\
&Y_{11}(q) = \frac{N q}{N + N_f - q}+ h^2 N_f\frac{N   (N q (N_f - q) + N_f + N )}{(N + N_f - q) (N + q)} + \nonumber \\
&+ h^4 \left[\frac{ N_f (N_f+1) }{2}\frac{(N-1) N (N_f - q) ( 2 N_f + 2N  + N q(N_f - q + 1) )}{2 (N + N_f - q) (N + q) ( N + q-1)} \right. + \nonumber \\
&\left.+ \frac{ N_f( N_f -1) }{2} \frac{N (N+1) (N_f - q)(2 N_f + 2N  + N q (N_f - q - 1))}{2 (N + N_f - q) (N + q) (N + q +1)} \right]+O(h^6) \nonumber \\
&Y_{12}(q)  = \frac{N (N_f-q)}{N+q}+h^2\frac{N N_f \left(N_f-q\right) \left(N_f (N (N+q)-1)-N q (N+q)\right)}{(N+q-1) (N+q) (N+q+1)}+ O(h^4) \nonumber
\\
&Y_{21}(q) = \frac{N q \left(N_f (N q-1)+N q (N-q)\right)}{2 \left(N_f+N-q-1\right)
   \left(N_f+N-q\right) \left(N_f+N-q+1\right)} + O (h^2) \nonumber \\
&Y_{22}(q) = \frac{N \left(N q \left(N_f-q\right)+N_f+N\right)}{(N+q) \left(N_f+N-q\right)} + O (h^2) \nonumber \\ 
&Y_{23}(q) = \frac{N \left(N_f-q\right) \left(N_f (N (N+q)-1)-N q (N+q)\right)}{2 (N+q-1) (N+q)
   (N+q+1)} + O (h^2) \nonumber
\end{eqnarray}

\noindent
{\bf Small mass expansion:}
\begin{align}
\label{sun_pf_small_mass}
&\Xi (g_{\pm},m\approx 0)=  \sum\limits_{q=0}^{N_f}  \ Z_{0}^{N,N_f}(q) \
\biggl [   H_0 (q) + N (g_+ H_1(-q)+ g_- H_1(q))+\nonumber  \\
& + N^2(g_+^2 H_2(q)+ g_+ g_- H_{12}(q)+ g_-^2 H_{2}(-q))+ O(g_{\pm}^3)\biggr] \ ,  
 \end{align}
where
\begin{eqnarray}
& Z_{0}^{N,N_f} (q)=  \underset{1\le k,l \le N}{\det} \ \binom{2 N_f}{N_f-k+l + q} =\frac{ G[N+ 2 N_f+1] G[N+1] G[N_f-q+1] G[N_f+q+1] }
{ G[2 N_f+1] G[N+ N_f- q+1] G[N+ N_f+ q+1] }\ . \\
\label{Z_0NP1}
&H_0 (q)=1 + \frac{s \ t}{\left(4N_f^2-1\right)}\left[\frac{m^{2}}{2!} (\frac{3 (s+2) (t-2)}{4 N_f^2-9}-2) + \right. \\
&+ \frac{m^4}{4!} (\frac{20 (s+2)  (t-2) (s+6)(t-6)}{\left(4 N_f^2-25\right) \left(4N_f^2-9\right)}
+ \frac{10 s t (s+2) (t-2)}{\left(4 N_f^2-9\right) \left(4 N_f^2-1\right)}  -   \nonumber  \\
& -\frac{15 s t (s+6)  (t-6)}{\left(4 N_f^2-25\right) \left(4 N_f^2-1\right)} - \frac{5 (s+6) (t-6)}{4 N_f^2-25} -\frac{45 (s+2) (t-2)}{4 N_f^2-9} +
\frac{20 s t}{4 N_f^2-1}+16) \left. + O(m^6) \right] \ , \nonumber \\
&H_1(q)=\frac{N \left(N_f+q\right) }{N_f+N-q } \left(1+ \frac{m^2}{2}\left(\frac{\left( t \, s -2 \left(2 N_f+N\right) \left(N_f-q\right) \right)}{4 N_f^2-1}+1\right)+ O(m^4) \right)
 \ , \nonumber \label{sun_pf_small_mass_coeff1} \\
&H_2(q)=\frac{N \left(N_f+q\right) \left(N^2(N_f+q) +t -2N_f)\right)}{2
   \left(N_f+N-q-1\right) \left(N_f+N-q\right) \left(N_f+N-q+1\right)}+ O( m^2)\ , \nonumber \\
&H_{12}(q) =\frac{N^2\,  t +s}{\left(N_f+N-q\right)
   \left(N_f+N+q\right)}-1+ O( m^2) \ , \nonumber \\
   &s = N (N + 2N_f ) \ , \ t= N_f^2- q^2 \nonumber 
\end{eqnarray}

Both  expansions were performed  up to $h^{18}$, $g^9$ and $m^{12}$ , $g^9$ orders correspondingly.

\subsection{Straight   expansions of free energies in the `t Hooft-Veneziano limit:}

\label{Strict}
{\bf   Case $g_+ = g_- \equiv g$}

\begin{eqnarray}\label{F2U(N)}
F_2 [\kappa, g, m] =  F_{g=0} [\kappa, m ]-\frac{4 g(\xi-1)}{\xi^2-1}- \nonumber \\
-(\xi-1) \left \{ \frac{G}{(\xi-1)^2(\xi+1)}-\frac{ G^2 \xi^2 }{2 (\xi-1)^2 (\xi+1)^4} +\frac{ G^3 \xi^2 }{3 (\xi+1)^7} -\right. 
 \nonumber \\
- \frac{ G^4 \xi^2 (\xi^2-4 \xi+1) }{4 (\xi+1)^{10}} +\frac{G^5 \xi^2 \left(\xi^4-10 \xi^3+22 \xi^2-10 \xi+1\right)}{5 (\xi+1)^{13}}-
 \nonumber \\
\left. - \frac{G^6 \xi^2 \left(\xi^6-18 \xi^5+88 x\xi^4-150 \xi^3+88 \xi^2-18 \xi+1\right)}{6 (\xi+1)^{16}}  + O(G^7) \right \}  \, .
\end{eqnarray}

where 
$$
F_{g=0} [\kappa, m ] = \frac{\xi \log \left(4 \cosh ^2\left(\frac{m}{2}\right)\right)}{\xi-1} -\frac{(\xi+1)^2 \log (\xi+1)-2 \xi^2 \log \xi+(\xi-1)^2 \log (\xi-1)}{2 (\xi-1)} 
$$
and $ G = \frac{g}{8} \cosh^2\left(\frac{m}{2}\right)  $, $\xi= 2 \kappa +1$ .

Generation functional for the free energy is moderately simple and gives us all this expansion: 
\begin{eqnarray}\label{crtwi}
     GF[ g, \kappa]= \frac{\partial}{\partial g }  F_2 [\kappa, g, h=1]-\frac{2}{\kappa } =  \frac{1}{g}\frac{2 \overline{g} \left(2 \overline{g}^2-3 \overline{g}+\kappa +1\right) }{ \kappa  (2 \overline{g}-2 \kappa -1)}
\end{eqnarray}
where $\overline{g}=\overline{g}(G, \kappa)$ is as in  (\ref{solUN}) and $m=0$ . 

And generation functional meets next functional equation 
$$GF[ g, \xi]= 8- GF[ g/\xi , 1/\xi]$$
and differential equation of third order:
\begin{eqnarray}
    g^3 \frac{\partial^3 GF}{\partial g^3} +3 g^2 (\xi -1) \frac{\partial^3 GF}{  \partial g^2 \partial \xi}  +3 g (\xi-1) \xi \frac{\partial^3 GF}{ \partial g\partial^2 \xi } +(\xi^2-1) \xi \frac{\partial^3 GF}{\partial \xi^3}+ \nonumber \\
    +6 \left( g^2 \frac{\partial^2 GF}{\partial g^2} +g (2 \xi-1)  \frac{\partial^2 GF}{ \partial \xi \partial g} +
   \xi^2  \frac{\partial^2 GF}{\partial \xi^2} + g \frac{ \partial GF}{\partial g}+ \xi\frac{\partial GF}{\partial \xi}\right) =0 \, .
\end{eqnarray}

{\bf  General case $g_+ \neq g_-$}

In case of $U(N)$ and $g_+ \neq g_-$ we have next expansion for both free energy (with re-summation by $m$):
$$
F_1 [\kappa, g_+, g_-, m] = - \log h   - \kappa \log(1- h^2) + (g_++ g_-) h + \frac{g_+ g_-}{ \kappa}  ,
$$
$$
F_2 [\kappa,  g_+, g_- , m]=  F_{g=0} [\kappa, m]+ \frac{1+2 k}{k} \log 4 M + \frac{(g_+ +g_-)}{ k}\left(1- \frac{M }{1+k} \right)+
$$
$$
+
(g_+^2 +g_-^2)\frac{(2 k+1)  ((2 k+1)M^2 -(1+ k)^2 M)}{4 k (k+1)^4 } +g_+ g_- \frac{(2 k+1) M }{k (k+1)^2} +
$$
$$
+\frac{\left(g_-^3+g_+^3\right) k (2 k+1) \left(3 (k+1)^2 M^2-4 (2 k+1)
   M^4\right)}{3 (k+1)^7}-\frac{\left(g_+ g_-^2+g_+^2 g_-\right) k (2 k+1)
   M^2}{(k+1)^5} +
$$
$$
+ O ( (g_+^4, g_-^4))
$$
where $M=\cosh ^2\frac{m}{2}$. 
The critical  line is calculated by equalizing $F_1$ to $F_2$ and expanding $h = e^{-m}$ in series by $g_+$, $g_-$ :
\begin{eqnarray}
  e^{ \mp m} = \frac{1 -  g_+ - g_-}{\xi -  g_+ - g_-} + \frac{2 (g_+ - g_-)^2 \kappa^3}{(1 + k) (1 + 2 k)^3}\left(1 +\frac{ (g_+ + g_-) (8 \kappa +5) }{(\kappa +1) (2 \kappa +1)} - \right. \nonumber \\
   -\frac{\left(g_-^2+g_+^2\right) (\kappa^4   +15\kappa^3 -31\kappa^2-48\kappa-15)-2 g_- g_+
   (\kappa^4   + 3\kappa^3 +41\kappa^2-48\kappa-15)}{(\kappa +1)^2 (2 \kappa +1)^2}+  \nonumber \\ 
   + \frac{ \left(g_+ g_-^2+g_+^2 g_-\right) \left(-96 \kappa
   ^4+352 \kappa ^3+795 \kappa ^2+504 \kappa +105\right)}{(\kappa +1)^3 (2 \kappa +1)^3} +  \nonumber \\ 
   + \frac{ \left(g_-^3+g_+^3\right) (  
  - 224 \kappa^4  - 96 \kappa^3 + 205\kappa^2 +168 \kappa+35)}{(\kappa +1)^3 (2 \kappa +1)^3}  + \left.O ( (g_+^4, g_-^4) \right)\ . \label{full_crit}
\end{eqnarray}
that guarantee 3rd order phase transition (as we understand from first terms of expansion). 
The solution  of the mean field  equation is the same as in the restricted case but formally includes the chemical potential.


\begin{thebibliography}{99}


%
\bibitem{Philipsen_19} O. Philipsen, PoS LATTICE2019 (2019) 273 [arXiv:1912.04827 [hep-lat]], O.~Philipsen, J.~Scheunert, JHEP {\bf 11} (2019) 022
[arXiv:1908.03136 [hep-lat]].
%
\bibitem{pl_dual20} O.~Borisenko, V.~Chelnokov, S.~Voloshyn, Phys.Rev. D {\bf 102} (2020) 014502 [arXiv:2005.11073 [hep-lat]].
%
\bibitem{Hooft_74} G.~t' Hooft, Nucl.Phys. B {\bf 72} (1974) 461.
%
\bibitem{Veneziano_76} G.~Veneziano, Nucl.Phys. B {\bf 117} (1976) 519.
%
\bibitem{gross_witten} D.~J.~Gross, E.~Witten, Phys.Rev. D {\bf 21} (1980) 446.
%
\bibitem{wadia} S.~R.~Wadia, Phys.Lett. B {\bf 93} (1980) 403.
%
\bibitem{damgaard_patkos} P.~H.~Damgaard and A.~Patk\'{o}s, Phys.Lett. B {\bf 172} (1986) 369.
%
\bibitem{christensen12} C.~H.~Christensen, Phys.Lett. B {\bf 714} (2012) 306
[arXiv:1204.2466 [hep-lat]].

%
\bibitem{pisarski18} H.~Nishimura,, R.~D.~Pisarski, V.~V.~Skokov,  Phys.Rev. D {\bf 97} (2018) 036014 [arXiv:1712.04465 [hep-th]].
%
\bibitem{pl_largeN_conf21} O.~Borisenko, V.~Chelnokov, S.~Voloshyn,  PoS LATTICE \textbf{2021} (2021) 453 [arXiv:2111.07103 [hep-lat]]
%
\bibitem{pl_largeN21} O.~Borisenko, V.~Chelnokov, S.~Voloshyn, Phys.Rev. D {\bf 105} (2022) 014501 [arXiv:2111.00474 [hep-lat]].
%
\bibitem{Russo_2020} Jorge G. Russo, Phases of unitary matrix models and lattice QCD2, arXiv:2010.02950v1
%
\bibitem{Russ_2021} Jorge G. Russo and Miguel Tierz,  Multiple phases in a generalized Gross-Witten-Wadia, arXiv:2007.08515v1 [hep-th]
%
\bibitem{SANTILLI_2020} L. Santilli, M. Tierz,  Exact equivalences and phase discrepancies between random matrix ensembles, arXiv:2003.10475v2 [math-ph]
%
\bibitem{1D_QCD} O. ~Borisenko, V.~Chelnokov, S.~Voloshyn, P.~Yefanov, One-dimensional QCD at finite density and its 't Hooft-Veneziano limit. JHEP 01 (2025) 008, [arXiv:2410.02328  [hep-lat]].
%
\bibitem{sun_int} O. Borisenko, S. Voloshyn, V. Chelnokov, Rep. Math. Phys. {\bf 85}  (2020) 129 [arXiv:1812.06069 [hep-lat]].

%
\bibitem{orthog_polynom} Y.Y.~Goldschmidt, 1/ N expansion in two-dimensional lattice gauge theory, J. Math. Phys. {\bf 21} (1980) 1842, DOI: 10.1063/1.524600.
%
\bibitem{un_dual18} O.~Borisenko, V.~Chelnokov, S.~Voloshyn, EPJ Web Conf. {\bf 175} (2018) 11021 [arXiv:1712.03064 [hep-lat]].
%
\bibitem{mcdual_21}
O. Borisenko, V. Chelnokov, E. Mendicelli, A. Papa, Nucl.Phys.B 965 (2021) 115332
[arXiv:2011.08285 [hep-lat]].
%
%
\bibitem{duals_abelian}
O.~Borisenko, V.~Chelnokov, S.~Voloshyn, P.~Yefanov,
Phys.Lett. B {\bf 827} (2022) 137000
[arXiv:2112.06002 [hep-lat]].

%
\bibitem{largeN_sun} O.~Borisenko, V.~Chelnokov, S.~Voloshyn,
Nucl.Phys. B960 (2020) 115177 [arXiv:2008.00773 [hep-lat]].




\end{thebibliography}
\end{document}